%% file: MN-17-1709-MJ.R3.tex
\documentclass[a4paper,fleqn,usenatbib]{mnras}
\usepackage[T1]{fontenc}
\usepackage{ae,aecompl}
\usepackage{graphicx}	
\usepackage{amsmath}	
\usepackage{amssymb}	
\usepackage{epstopdf}
\usepackage{gensymb}
\usepackage{txfonts}
\title[Properties of iodine absorption cells]{A systematic approach to determining the properties of an iodine absorption cell for high-precision radial velocity measurements}

\author[V. Perdelwitz et al.]{V. Perdelwitz$^{1}$\thanks{E-mail: vperdelwitz@hs.uni-hamburg.de}
P. Huke$^{2}$\thanks{E-mail: huke@astro.physik.uni-goettingen.de}
  \\
$^{1}$Hamburger Sternwarte, Universit\"at Hamburg, Gojenbergsweg 112, 21029 Hamburg, Germany\\
$^{2}$Institut f{\"u}r Astrophysik, Georg-August-Universit\"at G\"ottingen, Friedrich-Hund-Platz 1, G{\"o}ttingen, Germany}

\date{Accepted 2018 June 5. Received 2018 June 5; in original form 2017 May 20.}

\pubyear{2018}

\begin{document}
\label{firstpage}
\pagerange{\pageref{firstpage}--\pageref{lastpage}}
\maketitle

\begin{abstract}
Absorption cells filled with diatomic iodine are frequently employed as wavelength reference for high-precision stellar radial velocity determination due their long-term stability and low cost. Despite their wide-spread usage in the community, there is little documentation on how to determine the ideal operating temperature of an individual cell. We have developed a new approach to measuring the effective molecular temperature inside a gas absorption cell and searching for effects detrimental to a high precision wavelength reference, utilizing the Boltzmann distribution of relative line depths within absorption bands of single vibrational transitions. With a high resolution Fourier transform spectrometer, we took a series of 632 spectra at temperatures between 23~$\degree$C and 66~$\degree$C. 
These spectra provide a sufficient basis to test the algorithm and demonstrate the stability and repeatability of the temperature determination via molecular lines on a single iodine absorption cell. The achievable radial velocity precision $\sigma_{RV}$ is found to be independent of the cell temperature and a detailed analysis shows a wavelength dependency, which originates in the resolving power of the spectrometer in use and the signal-to-noise ratio. Two effects were found to cause apparent absolute shifts in radial velocity, a temperature-induced shift of the order of $\sim 1$~ms$^{-1}$K$^{-1}$ and a more significant effect resulting in abrupt jumps of $\geq 50$~ms$^{-1}$ is determined to be caused by the temperature crossing the dew point of the molecular iodine.

\end{abstract}


\input{Introduction.tex}

\input{Theory.tex}
\input{SetupExperiments.tex}

\input{Analysis.tex}

\input{Results.tex}

\input{Summary.tex}

\section*{Acknowledgements}

VP acknowledges funding through DLR and DFG.
PH acknowledges funding from the FP-7 starting grant agreement number 279347 and the DFG Heisenberg Grant RE1664/9-2.
We would like to thank Artie Hatzes of the Th\"uringer Landessternwarte Tautenburg for manufacturing the iodine cell.

\bibliographystyle{mnras}
\bibliography{Astro.bib} 
\label{lastpage}
\end{document}

%% file: Introduction.tex
\section{Introduction}
\label{sec:Introduction}

High-precision radial velocity (RV) measurements of stars are to date one of the most successful techniques for the discovery of exoplanets~\citep{seager2010exoplanets}. The required precision of the measurements can be on the order of $\sim1$~m/s, and necessitates, beside a high signal-to-noise ratio (SNR) of the data and resolving power of the spectrograph in use, a very precise knowledge of the wavelength scale. While the use of laser frequency combs \citep{2008Natur.452..610L} for this purpose is becoming more and more popular, the costs involved are too high for many observing facilities. 
Low cost alternatives such as Fabry-P\'erot etalons \citep[e.g.][]{2016arXiv160705172S}, hollow-cathode calibration lamps \citep[e.g.][]{1996A&AS..119..373B} and gas absorption cells \citep[e.g.][]{2002A&A...392..671E} are available. \\
Especially in the  optical regime, iodine cells present an affordable yet highly accurate alternative, since iodine exhibits a sufficiently high line density.  First introduced by \citet{Marcy1992}, iodine absorption cells have since been successfully employed 
in the search for exoplanets~\citep[e.g.][]{2016PhDT........74W,2010SPIE.7739E..4TC}.
\\
In order to guarantee a stable wavelength solution and obtain a high RV precision, the properties of an individual iodine cell have to be determined via high-resolution laboratory spectroscopy. Researching previous publications on the topic, one finds that, despite the frequent use of such reference cells, these laboratory measurements are to date not well documented. Most authors restrict the published information on mentioning the applied cell temperature and the method of heating.\\
In this publication we present a systematic approach to determining the properties of an iodine cell, such as absorption, stability and achievable radial velocity precision by using a high resolution Fourier Transform Spectrometer (FTS). Since the analyzed gas cell will be implemented in the robotic telescope {\it TIGRE} \citep{2014AN....335..787S} and therefore has to be fully compatible with autonomous operation, we also infer the minimum heating period required before the start of telescope operation each night in order to ensure a stable equilibrium inside the cell.\\
This publication is structured in the following way. In Section~\ref{sec:theory} we give an overview of the theoretical background of high-precision RV measurements and the thermodynamics of diatomic molecules. In Section~\ref{sec:setup} we describe the experimental setup. Section \ref{sec:analysis} details the analysis of the data acquired and in Section~\ref{sec:results} we present and discuss our results.

%% file: Theory.tex
\section{Theory}
\label{sec:theory}

\subsection{RV-precision with iodine reference cells}
The precision of any wavelength calibration (see e.g. \cite{2010ASSL..366...51H} and \cite{2010exop.book...27L} for an overview of the methods) is limited to the precision of the reference spectrum template. A measured spectrum can be modeled as the convolution of the instrumental point spread function (PSF) and the product of the stellar spectrum $I_S$ and the iodine cell transmission function $T_{I_2}$:
\begin{equation}
\begin{aligned}
 &I_{obs}(\lambda)=k\left[I_S(\lambda+\Delta \lambda_S)T_{I_2}(\lambda+\Delta \lambda_{I_2})\right]\ast PSF
\end{aligned}
\end{equation}
where $\ast$ denotes the convolution, $\Delta\lambda_S$ and $\Delta\lambda_{I_2}$ represent the RV shifts in the stellar and iodine spectra and k is a factor of proportionality. Following \cite{Marcy1992}, the final velocity shift can then be determined as 
\begin{equation}
\begin{aligned}
\Delta\lambda=\Delta\lambda_S-\Delta\lambda_{I_2}
\end{aligned}
\end{equation}
and hence the overall radial velocity precision is limited by the iodine cell calibration.\\
In order to derive a relationship between the achievable radial velocity precision and parameters of the spectrum, we fit all lines with a gaussian profile.
From the normalized gaussian with variance $\sigma$
\begin{equation}
\begin{aligned}
f(x)=\frac{1}{\sqrt{2\pi}\cdot\sigma}\text{exp}\left[-\frac{1}{2}\frac{(x-x_0)^2}{\sigma^2}\right]
\end{aligned}
\label{gaussianstandard}
\end{equation}
and using 
\begin{equation}
\begin{aligned}
f_{m}=2*f_{hm}=f\left(\pm\frac{\sigma_{FWHM}}{2}\right)
\end{aligned}
\end{equation}
where $f_{m}$/$f_{hm}$ denote the maximum/half maximum and $\sigma_{FWHM}$ is the full width at half maximum, it follows that
\begin{equation}
\begin{aligned}
\frac{1}{2}=\text{exp}\left[-\frac{1}{2}\frac{(x-x_0)^2}{\sigma^2}\right]
\end{aligned}
\end{equation}
which can then be solved for the locations of the half maxima $x_+$ and $x_-$ and thus yields
\begin{equation}
\begin{aligned}
\sigma_{FWHM}=x_+-x_-=2\sigma\sqrt{2\cdot\text{ln}2}.
\end{aligned}
\end{equation}
The gaussian line profile can then be descibed as
\begin{equation}
\begin{aligned}
n(\bar{\nu})=\frac{2}{\sigma_{FWHM}}\sqrt{\frac{\text{ln} 2}{\pi}}\text{exp}\left[-4\,\text{ln} 2\left(\frac{\bar{\nu}-\bar{\nu_i}}{\sigma_{FWHM}}\right)^2\right]
\end{aligned}
\label{gaussiandist}
\end{equation}
where
\begin{equation}
\begin{aligned}
\sigma_{FWHM}=\frac{\bar{\nu_i}}{c}\sqrt{\frac{8\,kT\,ln2}{m}}
\end{aligned}
\label{fwhmgaussian}
\end{equation}
with the Boltzmann constant $k$, temperature $T$, the mean frequency $\bar{\nu_i}$ and the molecular mass $m$. Since the iodine cell was evacuated during the manufacturing process and the pressure is sufficiently low, the effect of collisional broadening can be neglected in our approach.\\
The achievable precision of fitting a single line is
\begin{equation}
\begin{aligned}
\sigma_{RV}\propto\frac{\sqrt{\sigma_{FWHM}}}{C\cdot S/N}
\end{aligned}
\end{equation}
with the contrast\footnote{i.e. line depth divided by continuum level} $C$ and the signal-to-noise ratio $\frac{S}{N}$. Increasing the number of lines to $N_{tot}$ and using an average contrast $\overline{C}$, the overall precision becomes
\begin{equation}
\begin{aligned}
\sigma_{RV}\propto\frac{\sqrt{\sigma_{FWHM}}}{\overline{C}\cdot S/N\sqrt{N_{tot}}}
\end{aligned}
\label{sigmarv}
\end{equation}
Looking at the parameters of the iodine cell, three of the variables in Equation~\ref{sigmarv} depend on the temperature $T$. 
While a higher temperature does result in more detectable and deeper lines, it also causes line broadening and, eventually, saturation.\\
Experimentally, the resolution of the spectrograph limits the achievable RV precision as described in \cite{1992ESOC...40..275H} to
\begin{equation}
\begin{aligned}
\sigma_{RV}\approx 1.45\times10^9\cdot\left(\frac{S}{N}\right)^{-1}\cdot R^{-1}\cdot B^{-\frac{1}{2}} 
\end{aligned}
\label{hatzesrv}
\end{equation}
in $[m/s]$, where $S/N$ denotes the signal-to-noise ratio, R is the resolving power of the spectrograph and the wavelength coverage B (in $\mathring{A}$) is assumed to be proportional to the number of lines.

\subsection{Molecular gas temperature}
\subsubsection{Rovibrational transitions}
In this section we give a brief introduction to the theory of rotational-vibrational (rovibrational) spectra of diatomic molecules. 
Rotational-vibrational spectra exhibit characteristic bands caused by a simultaneous rotational and vibrational transition \citep{Gerstenkom1985}. In a first approximation the energy levels of a diatomic molecule can be described as the sum of a harmonic oscillator and a rigid rotator:
\begin{equation}
\begin{aligned}
 E(\nu,J) & =E_{vib}(\nu)+E_{rot}(\nu,J)\\
\end{aligned}
\label{eq:energy}
\end{equation}
where $\nu$ and $J$ denote the vibrational and rotational quantum numbers of the state.\\
Since the molecule is not a rigid rotator, and thus the internuclear potential depends on the energy of the state, we have to account for the nonlinearity of the Morse potential by expanding the second term via a Dunham expansion \citep{PhysRev.41.721}:

\begin{equation}
\begin{aligned}
 E_{rot}(\nu,J) & =\sum\limits_{j,k}Y_{k,l}(\nu+1/2)^k\left[J(J+1)\right]^l\\
\end{aligned}
\end{equation}
In order to determine the observable energy of a transition from $(\nu'',J'')$ to $(\nu',J')$, \cite{Gerstenkom1985} limited the number of terms to 4 and 6, respectively:
\begin{equation}
\begin{aligned}
 E(\nu'',J'')=&E_{\nu''}(0)+B_{\nu''}K-D_{\nu''}K^2+H_{\nu''}K^3\\
 E(\nu',J')=&E_{\nu'}(0)+B_{\nu'}K-D_{\nu'}K^2+H_{\nu'}K^3\\
&+L_{\nu'}K^4+M_{\nu'}K^5\\
\end{aligned}
\label{expansionGL}
\end{equation}
with $K=J(J+1)$. The coefficients $E_{\nu''}(0)$, $B_{\nu''}$, $D_{\nu''}$, $H_{\nu''}$, $E_{\nu'}(0)$, $B_{\nu'}$, $D_{\nu'}$, $H_{\nu'}$, $L_{\nu'}$ and $M_{\nu'}$ are then determined empirically.\\
Since for our purposes (i.e. temperature determination) it is only necessary to calculate the transition energy to a degree of precision that allows for an unequivocal determination of the quantum numbers of a given transition and thereby absorption line, we limit our approach to the expansion described above and use the catalogue of molecular parameters provided by \cite{Gerstenkom1985}.\\
The following transitions between energy levels are allowed: $\Delta \nu= \pm 1$, ${\Delta} J = \pm 1$ for rotational-vibrational and $\Delta\nu=0$, $\Delta J=\pm1$ for purely rotational transitions\footnote{Note that transitions with $\Delta\nu=\pm1$, $\Delta J=0$ are forbidden in most molecules.}.  Transitions with $J'-J''=-1$ for a given $\nu$ constitute the {\it P-branch} and those with $J'-J''=+1$ constitute the {\it R-branch}.\\
This results in spectral lines of wavelengths

\begin{equation}
\begin{aligned}
\bar{\nu}_P&=E(\nu'',J'')-E(\nu',J')\\
&=E_{\nu''}(0)+B_{\nu''}J(J-1)-D_{\nu''}\left(J(J-1)\right)^2\\
&+H_{\nu''}\left(J(J-1)\right)^3-E_{\nu'}(0)-B_{\nu'}J(J+1)\\
&+D_{\nu'}\left(J(J+1)\right)^2-H_{\nu'}\left(J(J+1)\right)^3\\
&-L_{\nu'}\left(J(J+1)\right)^4-M_{\nu'}\left(J(J+1)\right)^5
\end{aligned}
\label{pbranch}
\end{equation}

\noindent for the {\it P-branch} ($\Delta J=-1$) and

\begin{equation}
\begin{aligned}
\bar{\nu}_R&=E(\nu'',J'')-E(\nu',J')\\
&=E_{\nu''}(0)+B_{\nu''}(J+1)(J+2)\\
&-D_{\nu''}\left((J+1)(J+2)\right)^2+H_{\nu''}\left((J+1)(J+2)\right)^3\\
&-E_{\nu'}(0)-B_{\nu'}J(J+1)+D_{\nu'}\left(J(J+1)\right)^2\\
&-H_{\nu'}\left(J(J+1)\right)^3-L_{\nu'}\left(J(J+1)\right)^4\\
&-M_{\nu'}\left(J(J+1)\right)^5
\end{aligned}
\label{rbranch}
\end{equation}

\noindent for the {\it R-branch} ($\Delta J=+1$).
\subsubsection{Boltzmann distribution of rovibrational states}\label{boltzmannstatistics}
\begin{figure}
  \resizebox{\hsize}{!}{\includegraphics{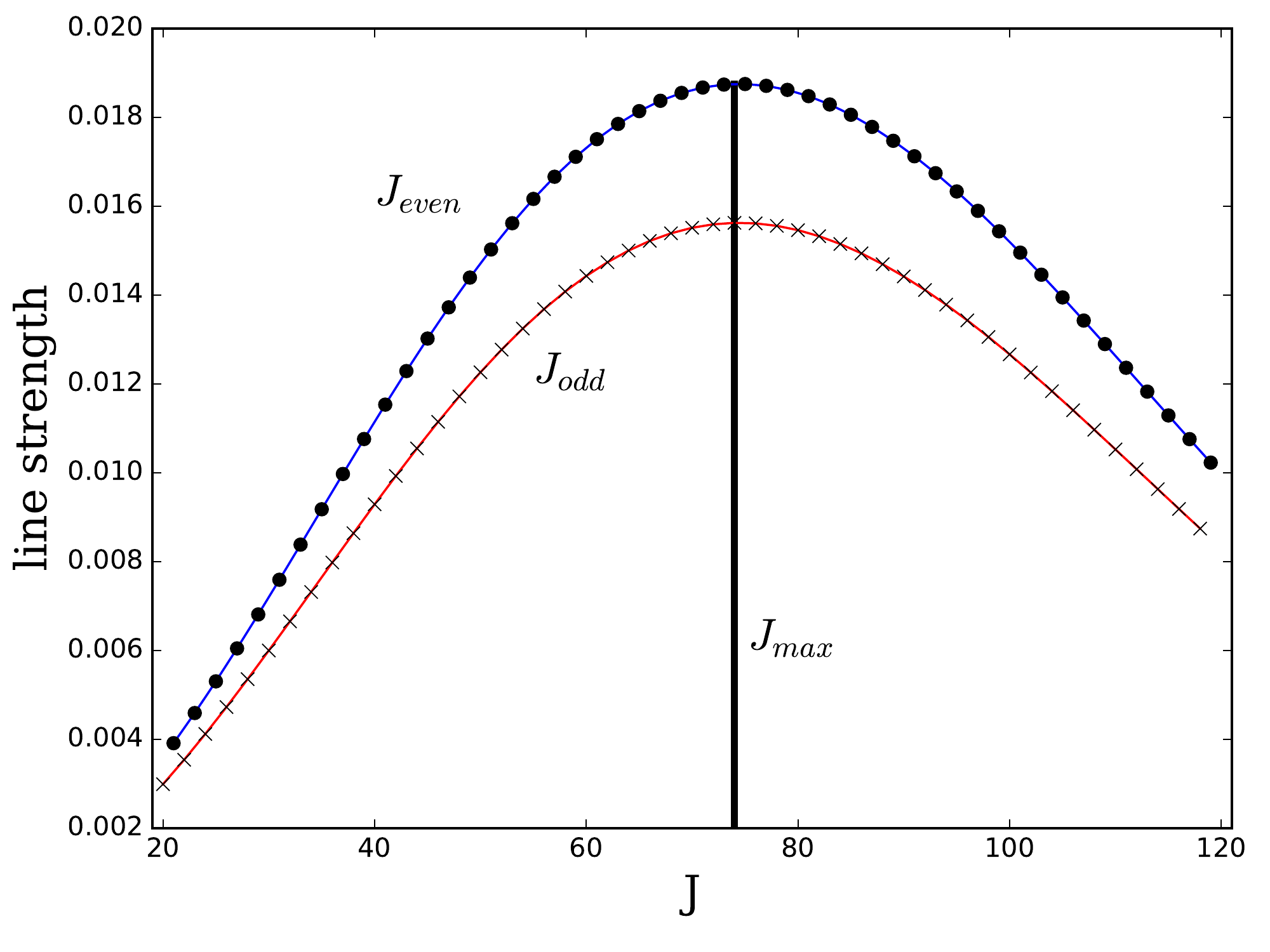}}
  \caption{Relative line strength versus rotational quantum number J in the R-branch of a single vibrational transition ($\nu''=12 \rightarrow \nu'=2 $). The upper curve shows transitions with even J, the lower curve with odd J. Since the maximum line strength is at the same value $J_{max}$ (vertical black line), both curves can be fitted simultaneously after multiplying the line strengths with odd J by a factor of $\frac{7}{5}$.}
  \label{fig:boltzmann}
\end{figure}

Equations \ref{pbranch} and \ref{rbranch} describe the splitting of each vibrational transition into different lines due to rotational transitions. The distribution of line strength within a given vibrational transition follows Boltzmann statistics:
\begin{equation}
\begin{aligned}
I_{rel}&=C_{\nu'',\nu'}\cdot g_{J''}\cdot S_{J'',J'}\cdot e^{-\Delta E_{J''}hc/kT}
\end{aligned}
\label{boltzmann}
\end{equation}
where $E_{J''}$ denotes the rotational energy, $g_{J''}$ is the degeneracy factor of $J''$ and $S_{J'',J'}$ is the two-photon transition strength \citep{1994JMoSp.163..541K}. While the physics behind the factors $g_{J''}$ and $S_{J'',J'}$ is interesting in itself, only the general dependency on $J$ is relevant for our purpose:
\begin{equation}
\begin{aligned}
\frac{I_{rel}}{I_0}&=(2J''+1)\cdot e^{-\Delta E_{J''}hc/kT}\\
&\approx(2J''+1)\cdot e^{-B_{\nu''}J''(J''+1)hc/kT}\\
\end{aligned}
\label{boltzmann2}
\end{equation}
Here, $\Delta E_{J''}$ is the energy of the state relative to the main vibrational transition.\\
The temperature inside the gas cell can be determined via the Boltzmann distribution given in equation \ref{boltzmann2}. Each vibrational wing of the spectrum exhibits a maximum at $J_{max}$ and setting the derivative of  Equation~\ref{boltzmann2} equal to zero yields
\begin{equation}
\begin{aligned}
T\approx \frac{hc}{2k}\cdot\Big[&-8D_{\nu''} J_{max}^4-16D_{\nu''} J_{max}^3+(4B_{\nu''}-10D_{\nu''}) J_{max}^2\\
&+(4B_{\nu''}-2D_{\nu''}) J_{max}+B_{\nu''}\Big]
\end{aligned}
\label{tempdet}
\end{equation}
where only the two dominant factors, $B_{\nu''}$ and $D_{\nu''}$, have been used in the approximation.\\
Since the iodine atoms are fermions with a nuclear spin of 5/2, the intensities of odd absorption lines are enlarged by a factor of 7/2 with regard to those of the even ones (see fig.~\ref{fig:boltzmann}). A good description of this effect can be found in \cite{1994JMoSp.163..541K}.

%% file: SetupExperiments.tex
\section{Setup and Data Acquisition}
\label{sec:setup}
\begin{figure}
  \resizebox{\hsize}{!}{\includegraphics{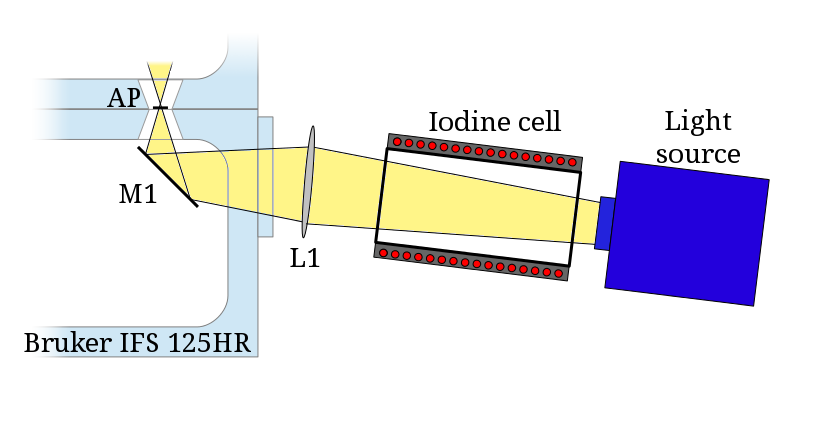}}
  \caption{Schematic setup of the experiment. Uncollimated light from the source passes through the iodine cell and is focused by lens L1 onto the FTS entrance aperture AP. Note that fold mirror M1 and the entrance aperture are inside the Bruker IFS 125HR.}
  \label{fig:setup}
\end{figure}

The analyzed iodine cell was manufactured and filled by A. Hatzes at the Th\"uringer Landessternwarte Tautenburg and equipped with a heating foil and temperature control electronics at the electronics workshop of the Hamburger Sternwarte\footnote{The cell filling process is the same as described in \cite{Cochran1994}.}. During the manufacturing process, the cell was evacuated and filled with iodine while heated at a temperature of $\approx30\degree$C, since it was originally supposed to be operated at a temperature between 50 and 70 degrees. This means that at a lower temperature the iodine will condense out, necessitating a very stable temperature control.\\
Measurements were carried out with a Bruker IFS 125HR, which is a commercial, high-resolution Fourier Transform Spectrograph (FTS) designed for laboratory applications. A detailed description of the instrument can be found in \cite{2016A&A...587A..65R}.\\
The iodine cell was placed in front of the entrance port of the FTS and a lens was used to focus the light onto the entrance aperture inside the FTS (see Figure~\ref{fig:setup}). A tungsten lamp was used as an external light source.

\noindent A series of measurements were carried out with the Bruker IFS 125HR in G\"ottingen on March 26-27 2015 in order to obtain a template spectrum of the iodine cell, test our temperature determination algorithm and determine the optimum cell temperature for stellar spectroscopy. Since the wavelength regime of interest is in the optical, a Si detector and Quartz VIS beam splitter were used in all runs. All spectra were taken in single double-sided scan mode, i.e. with the scanning mirror moving back and forth once per spectral acquisition, with an entrance aperture of 1~mm.\\
Preliminary measurements showed that the iodine content within the cell is comparatively high, resulting in a relatively low optimal cell temperature for use in stellar observations. For this reason the major part of the measurements were taken during one long, continuous run, during which the cell heating was switched on (with a nominal temperature setting of 50~$\degree$C) for a duration of two hours and subsequently turned off for one hour. This process was repeated five times and spectra were acquired throughout the process. Subsequently, 45 spectra were taken in the temperature range of 50 to 66~$\degree$C. In this way both the spectral response to different temperatures and the heating and cooling phases could be determined.\\
A high temporal resolution was required throughout the measurement in order to resolve changes in the spectra during heating and cooling phases. All spectra in this series were acquired with a resolution of $0.02$~cm$^{-1}$, as the resolving power of an FTS is proportional to the maximum optical path difference and hence acquisition time. Note that the coherence length does not play any role in this case, since the interferometer geometry of the FTS acts as a narrow-band filter.\\
Additionally, 42 spectra were acquired at a setting of 30~$\degree$C with a resolution of $0.01$~cm$^{-1}$ to obtain high-resolution template spectra for the RV determination via cross-correlation.

%% file: Analysis.tex
\section{Analysis}
\label{sec:analysis}

\subsection{Spectra Preprocessing}
\label{sec:spectra-preprocessing}
The raw spectral data obtained with the FTS were translated from the native {\it Opus}\footnote{\url{https://www.bruker.com/products/infrared-near-infrared-and-raman-spectroscopy/opus-spectroscopy-software.html}} data format to a simple ascii table with a custom python code and the wavelength was converted from wavenumbers ($cm^{-1}$) to $nm$. The remaining steps of the spectral processing were performed with {\it GNU Octave\footnote{\url{https://www.gnu.org/software/octave/index.html}}}.\\
The spectral range was reduced to the range where iodine lines are present (500~nm to 625~nm) and all spectra were then normalized by fitting the continuum with a 4th order polynomial and subsequent division. 
In order to identify absorption lines the algorithm searched for local minima and rejected lines with a depth $\leq 10\%$ as well as obvious blends with a limit of $0.02$~nm for proximity. A Gaussian was fitted to all remaining lines, and a cut-off at a $95\%$ $R^2$ fit confidence yielded a list of $\approx10000$ lines over the entire spectral range, each with a value for relative depth, FWHM and confidence level. Here, the $R^2$ fit confidence is given by
\begin{equation}
\begin{aligned}
R^2=1-\frac{\sum_{i=1}^n (y_i-\hat{y}_i)^2}{\sum_{i=1}^n (y_i-\bar{y})^2},
\end{aligned}
\label{rv_formula}
\end{equation}
where $y_i$ are the data and $\hat{y}_i$, $\bar{y}$ are the fit values and mean, respectively.

\subsection{Wavelength shift}
\label{sec:rv-drift}
The following steps were carried out with a custom python code utilizing the {\it PyAstronomy} package \footnote{\url{https://github.com/sczesla/PyAstronomy}} written by S. Czesla. Due to the fact that the iodine cell is used for radial velocity determination, we determine wavelength shifts in the same manner as one would when measuring stellar radial velocities, i.e. determine an effective RV-shift by properly Doppler-shifting the wavelength axis 
\begin{equation}
\begin{aligned}
\lambda_{eff}=\lambda_0\cdot\left(1+\frac{v_{eff}}{c}\right)
\end{aligned}
\label{rv_formula}
\end{equation}
where $v_{eff}$ is the best-fit effective RV-shift.\\
Due to the large file sizes a cross-correlation of the spectra with a sufficiently small step size was impossible. We therefor devised an iterative cross-correlation approach. The calculation was first performed choosing a maximum drift of $1$~km/s and a step size of $0.1$~km/s. The next window was chosen around the determined maximum, with a width of 1/10 of the previous window plus 1/100 on each side to account for the fact that the maximum might lie exactly at the edge of the window. This process is repeated 6 times (step size of 1~mm/s), after which the maximum of the curve was fitted with a gaussian, yielding a precise RV-shift value and error estimate. In order to check the validity of this approach, a cross-correlation of the full spectrum was carried out for a small sample.\\ 
A high signal-to-noise spectrum was computed as a template for the cross-correlation in the following manner: {\it (i)}~Determining the RV-shift for all 42 spectra acquired at 30~$\degree$C by performing a cross-correlation with the spectrum of highest SNR. {\it (ii)}~Correcting for the RV-shift. {\it (iii)}~Co-adding the corrected spectra.\\
Cross-correlation was then performed on all spectra of the sample, yielding an RV-shift and error estimate.

\subsection{Temperature determination}
\label{sec:temperature-gauge}
In this section we describe the temperature determination based purely on the spectra. The derived temperatures will be compared to the sensor measurements in Section~\ref{sec:transmissivity}.
In order to determine the temperature via Boltzmann statistics as described in Section~\ref{boltzmannstatistics}, $\sim10000$ lines were identified regarding their quantum numbers $\nu'',\nu'$~and~$J$. To do this an array of artificial wavelengths was produced using Equations~\ref{pbranch} \& \ref{rbranch} and the molecular constants given in table~V of \cite{Gerstenkom1985}. For all transitions $\nu''\rightarrow\nu'$ in the ranges $0<\nu''<80$ and $0<\nu'<\nu''$ the corresponding transition wavelengths for all values of J were determined and cross-matched with the spectral line list. The line strengths of even values of $J$ were multiplied with a factor of 7/5 to account for the different parities and allow for a simultaneous fit of even and odd values. If the number of close matches ($\delta\lambda\leq0.5\cdot\sigma_G$) exceeded the empirically determined required number of 10, a Boltzmann distribution (equation \ref{boltzmann2}) was fitted to the line strength as a function of $J$. 
After this, obvious outliers deviating by more than $3\sigma$ from the fit were rejected and a second fit, weighted by the discrepancy between preliminary fit and data, was performed. Since not all fits were successful due to unidentified blended lines, a threshold for the goodness-of-fit was then applied. The result was a list of Boltzmann curves for both the Q- and P-branch of each vibrational transition along with their standard errors. The exponential constant of each fit was then used to calculate the molecular temperature via Equation~\ref{tempdet}. Errors were determined from the errors of the single Boltzmann fits $\Delta T_i$ via
\begin{equation}
\begin{aligned}
\Delta T=\left(\sum_i{\frac{1}{\Delta T_i^2}}\right)^{-1/2}.
\end{aligned}
\end{equation}
\\

%% file: Results.tex
\section{Results and discussion}
\label{sec:results}
\subsection{Line detection}
\label{sec:linedetection}
\begin{figure}
  \resizebox{\hsize}{!}{\includegraphics{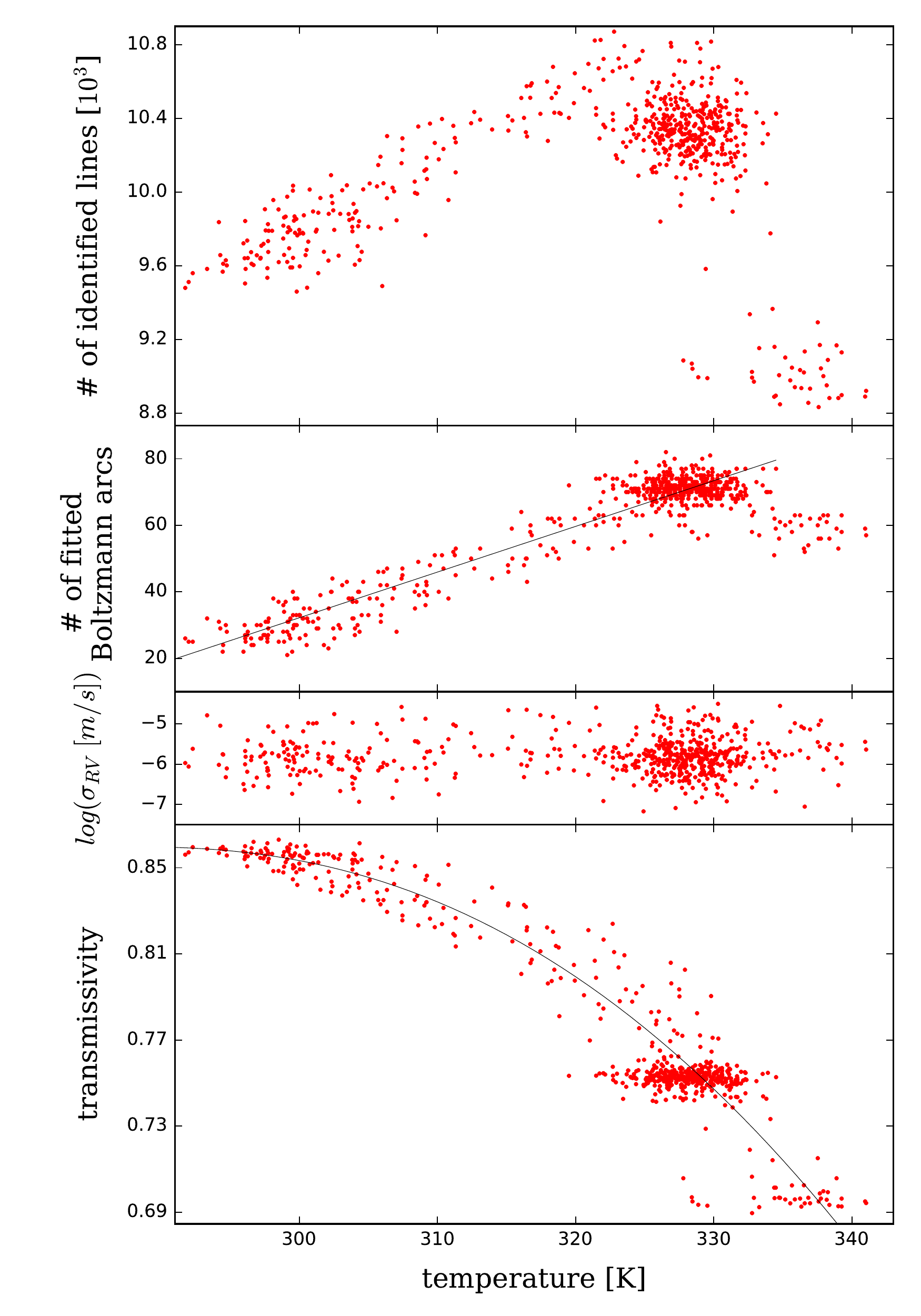}}
  \caption{Temperature-dependent properties of the iodine cell for the entire temperature range between ambient and 66~$\degree$C.  Temperatures have been determined with the algorithm described in Section~\ref{sec:temperature-gauge}. {\it Row 1:} Number of lines identified by the algorithm as a function of temperature. {\it Row 2:} Number of successful fits of Boltzmann distributions of single vibrational transitions as a function of cell temperature along with a best fit linear function describing the increase up to 325~K (black line). {\it Row 3:} Logarithm of the RV-precision in m/s as a function of temperature. There is no correlation between the achievable precision and cell temperature. {\it Row 4:}Total absorption of light as a function of temperature along with the best fit (solid line).}
  \label{fig:multiplotT}
\end{figure}

The total number of identified lines within our acceptance range (i.e. with a line depth of $d_{line}>0.1$) increases linearly up to a temperature of $325$~K and subsequently decreases (row 1 of Figure~\ref{fig:multiplotT}). We attribute the linear increase to shallow lines crossing the cut-off depth of $10\%$ with increasing temperature. The decrease at higher temperatures can be explained by more and more lines becoming saturated at higher temperatures, i.e. the quality of the fits of deep, saturated lines decreases, since they do not resemble gaussian profiles any longer (see Section~\ref{sec:spectra-preprocessing}).
We identify the temperature of $325$~K to be the point at threshold at which all iodine has transitioned to a gaseous state.
The number of successful fits of Boltzmann distributions of single vibrational transitions increases linearly until this temperature (Figure~\ref{fig:multiplotT}, 2nd panel), and thereafter decreases.\\

\subsection{Transmissivity}
\label{sec:transmissivity}
\begin{figure}
  \resizebox{\hsize}{!}{\includegraphics{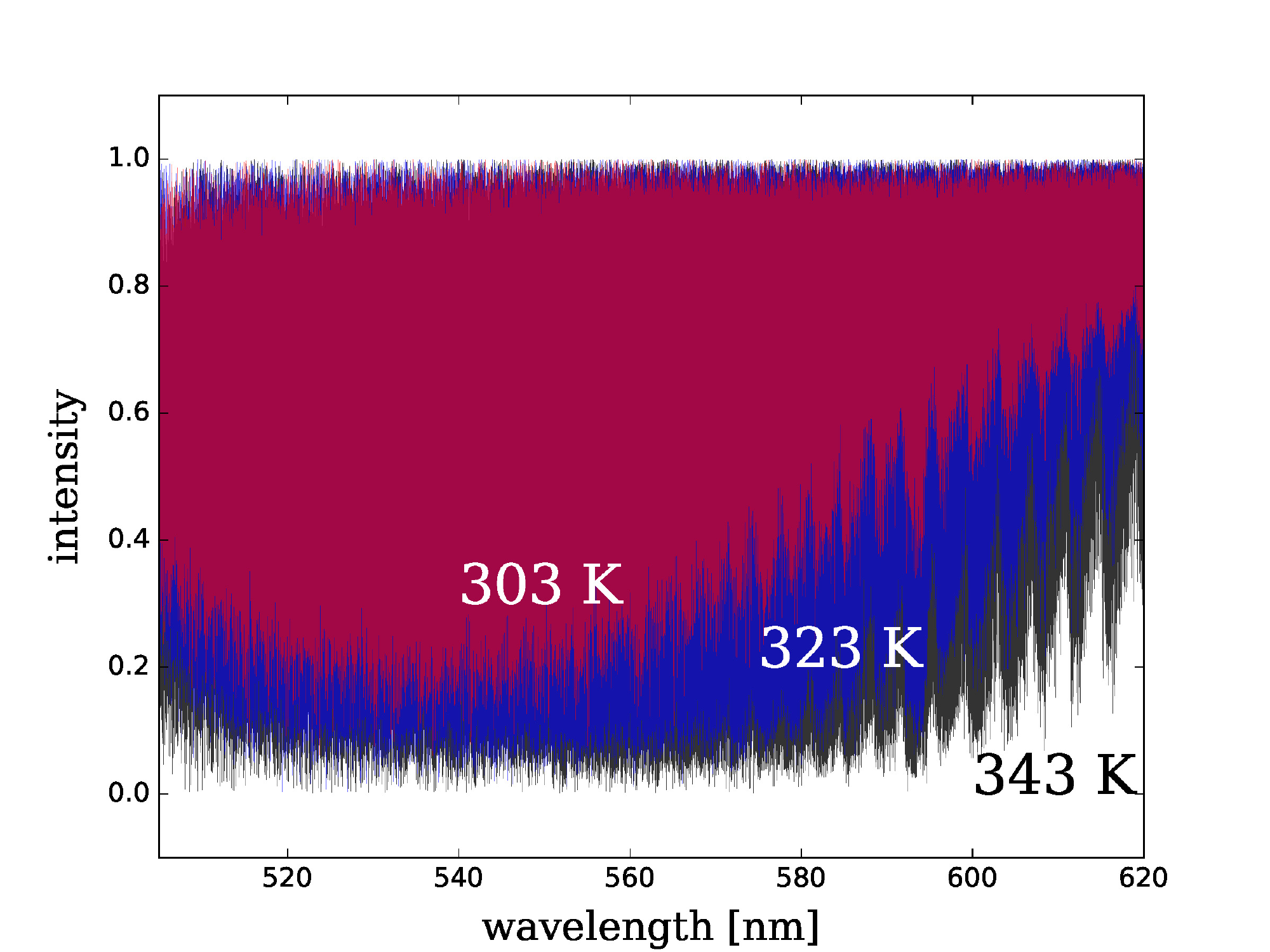}}
  \caption{Total spectrum for three different cell temperatures (303 K, 323 K, 343 K). At 323 K a large part of the spectrum is already saturated.}
  \label{fig:T_absorption}
\end{figure}

As already seen in row 1 of Figure~\ref{fig:multiplotT}, the number of identified lines increases with temperature as the depth of lines increases. 
Figure \ref{fig:T_absorption} shows the complete spectra for three different temperatures. In order to quantify the increase of absorption with temperature we numerically integrated all spectra. The resulting transmissivity can be seen in row 4 of Figure~\ref{fig:multiplotT} and confirms the expectation that a lower temperature yields a higher optical throughput of the source flux. 
In order to quantify the effect we fitted the data with a function
\begin{equation}
\begin{aligned}
\tau=\tau_0\cdot\left[1-\left(\frac{T-T_{0}}{b}\right)^d\right]
\end{aligned} 
\label{transmissivity}
\end{equation}
with proportionality constant $b$ and exponent $d$, zero absorption $\tau_0$ and temperature offset $T_0$. 
The best fit (black line in figure~\ref{fig:multiplotT}, bottom panel) yields a zero absorption of $\tau_0=0.86$, which we interpret as the maximum transmissivity of the cell, meaning that 12\% of the incoming flux is losses in the cell (e.g. via absorption or reflection). The temperature $T_0=288$~K can be interpreted as the threshold temperature beyond which the transmissivity decreases. The best fit proportionality constant and exponent are $b=105.2$ K$^{-1}$ and $d=2.2$.
For applications in stellar spectroscopy this iodine cell should thus be kept at a low temperature ($\approx 300$~K), where transmissivity is still in the range of $85\%$.\\ 
As can be seen in Figures~\ref{fig:multiplotT} and \ref{fig:T_absorption}, the spectrum at 303~K cell temperature already exhibits a large number of absorption lines covering the entire spectral range. In cells with a smaller amount of iodine the number of sufficiently deep lines will be lower at the red part of the spectrum, and cell temperature should be adjusted accordingly, until the entire range of interest is covered.\\

\subsection{Temperature determination}
\begin{figure}
  \resizebox{0.92\hsize}{!}{\includegraphics{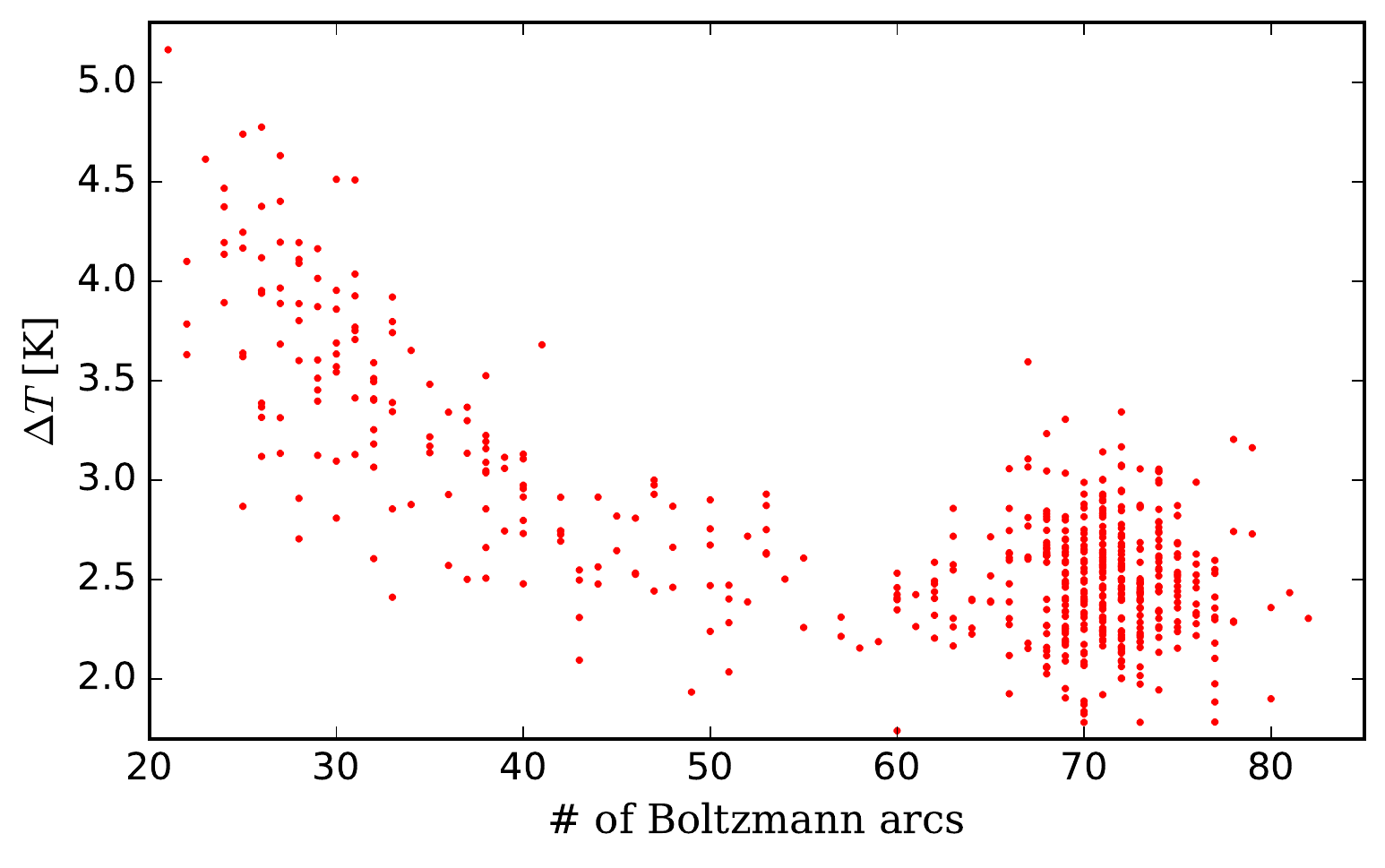}}
  \caption{Error in temperature determination as a function of the number of successful fits of Boltzmann distributions of single vibrational transitions.}
  \label{fig:nJ_deltatemp}
\end{figure}
The temperature determination as described in Section~\ref{sec:analysis} worked for all but a few spectra and yielded errors of $\leq 3$~K. As shown in Figure~\ref{fig:nJ_deltatemp}, the precision depends on the number of successful fits of Boltzmann distributions, while converging towards a constant value of $\approx2.5$~K. This maximum achievable precision can be explained by the temperature gradient inside the cell, which is expected to be fairly large due to the diameter of $6$~cm.\\
Row 1 of Figure~\ref{fig:amb-50} shows the temperature measured by the internal sensor (solid line) and determined by our algorithm (data points). There is a linear offset to the sensor temperature, the molecular temperature being systematically higher. This discrepancy can be corrected by a linear function
\begin{equation}
\begin{aligned}
T_{alg}=(T_{sensor}-T_{off,1})\times a + T_{off,2}
\end{aligned}
\label{temperatureoffset}
\end{equation}
A least-squares fit yields $a=1.414$, $T_{off,1}=292.2$~K and $T_{off,2}=290.45$~K. The dashed line in row~1 of Figure~\ref{fig:amb-50} shows the corrected sensor temperature. The residuals between corrected sensor temperature and molecular temperature ($\sigma=T_{measured}-T_{alg}$) displayed in row~2 show minor systematics during the unheated phases. The real cell temperature can then be determined from the algorithm output by inverting Equation~\ref{temperatureoffset}.\\
We consider three cases for the origin of the temperature offset and discuss them in the following sections.
\subsubsection{Erroneous model}
We assume the energy levels to include two terms, the vibrational and rotational energy (Equation~\ref{eq:energy}). In reality, there are several other contributors to the energy levels, such as a translational and an electronic term. We do, however, argue, that the omission of these terms cannot lead to a bias in temperature determination, since the temperature determination is only dependent on the difference between energy levels  $\Delta$E, hence any term independent of the quantum numbers $\nu$ and $J$ is canceled out in Equation~\ref{boltzmann2}.
\subsubsection{Heating by the source lamp}
The source lamp in our setup is located relatively close to the iodine cell, so heating of the gas cannot be dismissed outright. As can be seen in the lower panel of Figure~\ref{fig:multiplotT}, the absorption is in fact temperature dependent. The lamp therefore heats the gas and causes a temperature gradient between the cell center and the cell wall, where the temperature sensor is located.
But if this effect were large enough to be present in the data, the temperature offset would not drop to zero at lower temperatures (see upper panel of Figure~\ref{fig:amb-50} and Equation~\ref{temperatureoffset}), which is why we can rule it out as the underlying cause. 
\subsubsection{Heat sensor}
Equation~\ref{temperatureoffset} shows that the difference between the sensor and algorithm temperatures drops to zero at a temperature of 290.45~K, which was in fact the room temperature inside the laboratory at the time of the measurements. Furthermore, the residuals between corrected sensor temperature and the one determined via our algorithm shows systematics in the cooling phases, indicating that the sensor temperature drops faster when the heating is switched off.\\
These two observations lead us to conclude that the origin of the temperature deviation is an improper insulation of the temperature sensor, which loses heat either to the air or via the attached copper wire.

\subsection{RV-precision}

\begin{figure}
  \resizebox{\hsize}{!}{\includegraphics{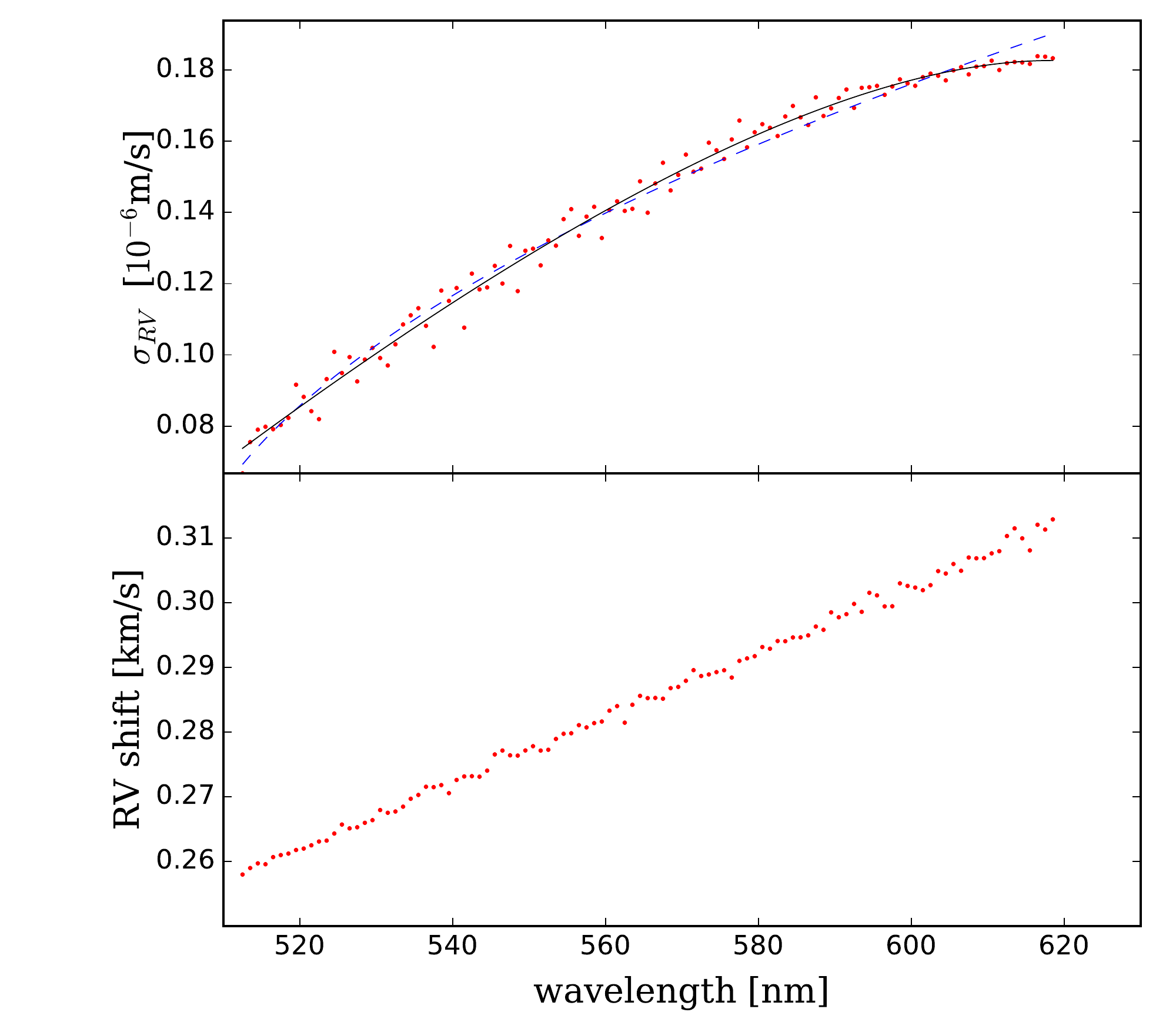}}
  \caption{Top row: Wavelength dependency of the RV precision $\sigma_{RV}$ for the combined spectra in bins of width 1~nm. The dashed blue line represents a square root law fit, while the solid black line has represents a logarithm multiplied with a linear function (Equation~\ref{logfunc}). Second row: Total effective RV/wavelength shift as a function of wavelength.}
  \label{fig:deltaRVandRV_temp}
\end{figure}

Aside from testing the new temperature determination algorithm, the main goal of our measurements was the determination of the ideal temperature setting for the iodine cell, particularly with regard to RV precision. It increases with the number of detected lines as $\sqrt{N}$ (see Equation~\ref{sigmarv}) and decreases with line width, which in turn increases with temperature as $\sqrt{T}$ (see equations~\ref{sigmarv} and \ref{fwhmgaussian}), yielding 
\begin{equation}
\begin{aligned}
\sigma_{RV}\propto\frac{T^{1/4}}{{N_{tot}}^{1/2}}.
\end{aligned}
\label{rvmax}
\end{equation}
The measured linear increase of the number of detected lines up to $\approx 325$~K and subsequent decline thus leads us to expect $\sigma_{RV}\propto T^{-1/2}$ up to a minimum and increase thereafter. 
We do, however, see no such dependency on temperature in the RV precision (see row 3 of Figure~\ref{fig:multiplotT}).\\
In order to test the dependency of the RV precision on other parameters, split the spectra into bins of width 1~nm, carry out the cross-correlation and compute the mean for of all spectra for each bin (top panel of Figure~\ref{fig:deltaRVandRV_temp}), since all spectra exhibit a similar wavelength dependency.
\begin{figure}
  \resizebox{\hsize}{!}{\includegraphics{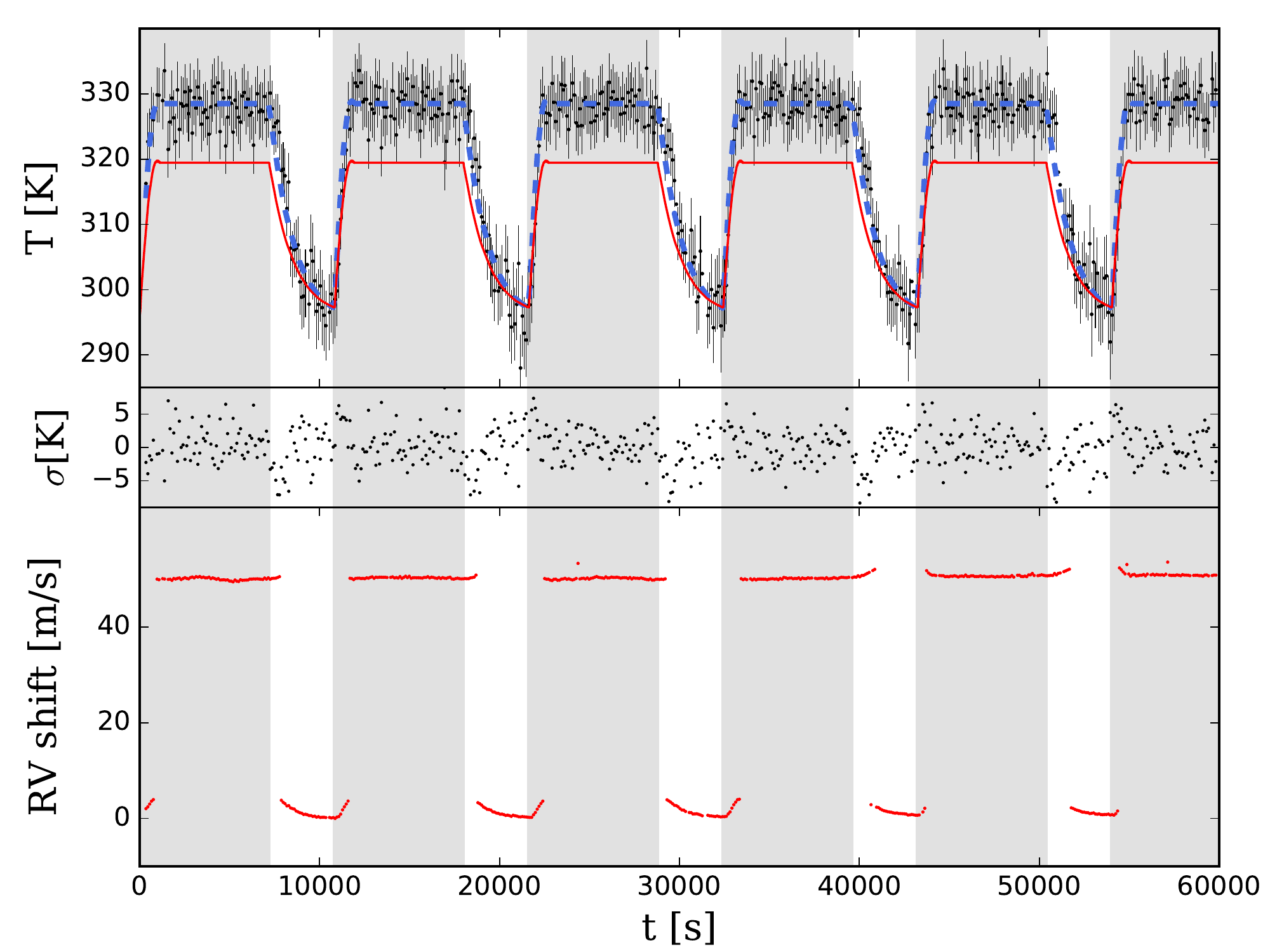}}
  \caption{Determined parameters of the measurements as a function of time. Shaded gray areas mark the time intervals during which the heating was switched on. Row 1: Data points with error bars represent the temperature of each spectrum as derived by our algorithm. The solid line shows the temperature determined with the gas cell sensor and the dashed line is the linearly corrected sensor temperature. Residuals between the corrected sensor temperature and the molecular temperature are given in row 2. Row 3 shows the determined effective RV/wavelength shift of each spectrum.}
  \label{fig:amb-50}
\end{figure}
We fit the data with a square root law (dashed blue line) and the product of a logarithm and a linear function (solid black line): 
\begin{equation}
\begin{aligned}
\sigma_{RV}=a_0\cdot\text{log}\left(a_1\lambda+a_2\right)*\lambda+a_3.
\end{aligned}
\label{logfunc}
\end{equation}
While the square root law describes the general trend of the data, the logarithmic function is a better fit. 
Since the spectral resolving power of the FTS \citep[e.g.][]{6953030} is given by
\begin{equation}
\begin{aligned}
R=\frac{\lambda^2}{2\cdot\delta_{max}},
\end{aligned}
\label{fts_res}
\end{equation}
where $\delta_{max}$ denotes the maximum displacement of the scanning mirror. Applying Equation~\ref{hatzesrv} then yields
\begin{equation}
\begin{aligned}
\left(\frac{S}{N}\right)\cdot\sqrt{B}\propto\frac{\lambda}{\text{log}(a_1\lambda+a_2)}
\end{aligned}
\label{fts_res}
\end{equation}
for the wavelength dependency of $S/N$ and B. Since the bandpass does not depend on temperature, this indicates that the signal-to-noise ratio is the limiting factor for achievable RV-precision.

\subsection{Wavelength shift}
Our data show a variation in the absolute wavelength shift during the measurements (see third row in Figure~\ref{fig:amb-50}). In the following sections we will present different mechanisms which are known to cause wavelength shifts and discuss whether they can contribute to the absolute wavelength shift.

\subsubsection{Temperature}
\label{sec:temperaturervdrift}
\begin{figure}
  \resizebox{\hsize}{!}{\includegraphics{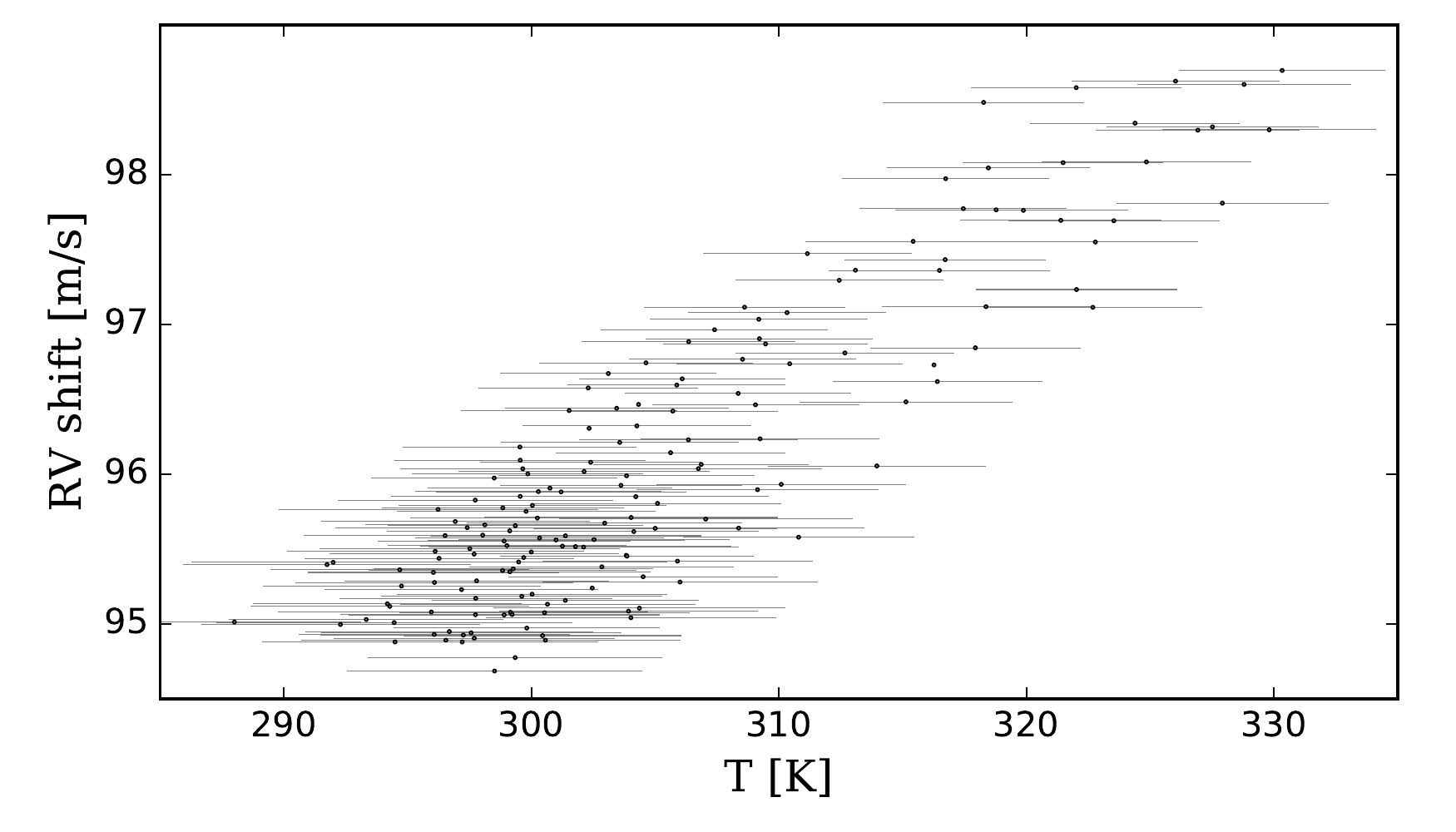}}
  \caption{Effective RV/wavelength shift as a function of temperature during periods where the heating is turned off.}
  \label{fig:T_RV}
\end{figure}

When the cell heating is turned off, the temperature gradually adjusts to the surroundings (see top row of Figure~\ref{fig:amb-50}), so that the influence of a temperature-dependent RV drift can be studied.\\
The wavelength shift during these periods (Figure~\ref{fig:T_RV}) shows a clear linear correlation with cell temperature. The absolute radial velocity of a spectrum is linearly dependent on pressure changes
\begin{equation}
\begin{aligned}
v_c=v_c^0+\frac{\delta P}{P_0}
\end{aligned}
\label{pressuredependence}
\end{equation}
where $v_c$ and $v_c^0$ are, respectively, the radial velocity and radial velocity at 30~$\degree$C and $\delta P$/$P_0$ is the relative change in pressure. Even though the amount of iodine in a gaseous state is increasing with temperature, the distribution of the wavelength shift indicates that $\delta P\propto \delta T$.

\begin{figure}
  \resizebox{\hsize}{!}{\includegraphics{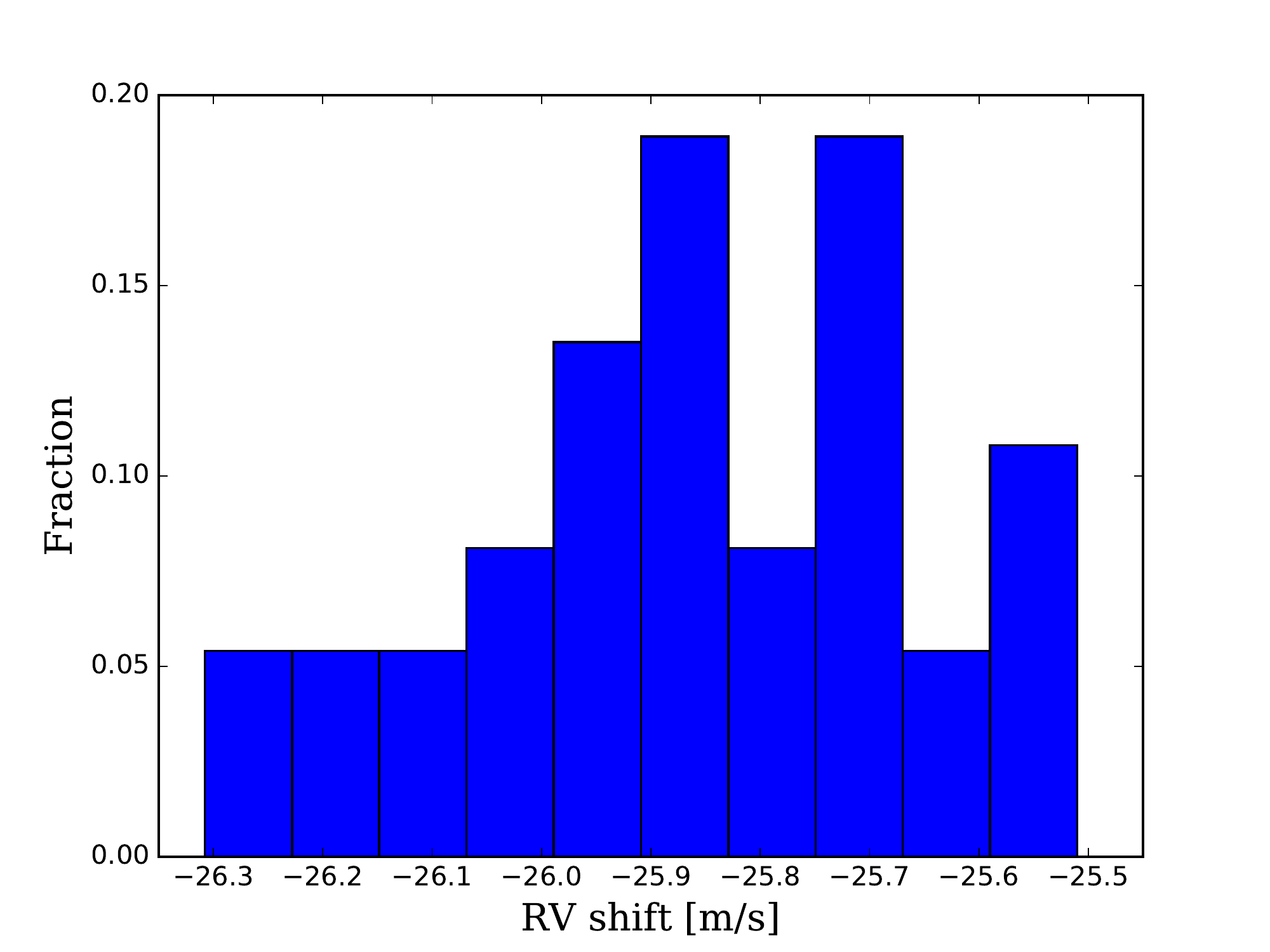}}
  \caption{Distribution of effective RV/wavelength shift in the sample of spectra at 303 K.}
  \label{fig:30degRVacc}
\end{figure}

The temperature-induced wavelength shift has a gradient of 0.11~$m\cdot s^{-1}\cdot K^{-1}$. We can compare this value to the distribution of wavelength shifts in the spectra acquired at a cell temperature of 30~$\degree$C (Figure~\ref{fig:30degRVacc}), which exhibits a standard deviation of 0.206~m/s, a value far below the RV precision of single spectra. Converting this into an uncertainty in temperature using the gradient mentioned above we obtain $\Delta T=1.9$~K, which is comparable to the mean errors of the temperature determination of 2.7~K.

\noindent Since the effect of a temperature-induced wavelength shift is of the order of $cm\cdot s^{-1}\cdot K^{-1}$, it can be controlled with a suitably precise heating mechanism. 

\subsubsection{Condensation/dew point}
As discussed in Section~\ref{sec:temperaturervdrift}, changes in pressure induce a radial velocity offset. Since the jumps in radial velocity (Figure~\ref{fig:amb-50}, bottom panel) occur at roughly the same temperature ($\sim325$~K), we conclude that they are caused by the crossing of the dew point of the iodine. This is supported by the fact that the transition temperature coincides with the point at which the number of detected lines does not increase any further, as described in section~\ref{sec:linedetection}. As explained in Section~\ref{sec:setup}, the iodine cell was filled in such a manner, that a low temperature will cause condensation of the iodine. In this case the delay between the start/stop of heating phases and the RV jumps would be caused by the inertia of heat propagation inside the glass windows.

%% file: Summary.tex
\section{Summary and Conclusion}
\label{sec:summary}
We have shown that fitting of Boltzmann distributions of single vibrational bands is a stable means of temperature determination in iodine reference cells. It yields the true molecular temperature and is not subject to the same systematic errors encountered by temperature sensors, which are usually situated close to the heat source, e.g. heating foil or water bath. Additionally, the method will allow for a more detailed study of temperature properties of gas reference cells, as it enables to probe layers of different distance relative to the center and wall of the cell with the use of narrow, collimated beams, thereby providing a measure of the temperature gradient, and to investigate whether a smaller cell diameter, and hence temperature gradient, improves the cell's performance. Furthermore, we have determined that the offset between the determined sensor temperature and that based on the Boltzmann distribution is likely caused by a problem with the temperature sensor.\\ 
There is a small dependency of the absolute RV drift on cell temperature, necessitating the cell heating to be stable to $<1$~K for high precision measurements. We argue that the large ($\approx50$~m/s) RV jumps and the maximum of detected lines originate in the temperature crossing the dew point of molecular iodine. It is important to avoid operating this particular iodine cell at temperatures in the vicinity of $\sim325$~K so as to prevent said jumps within single measurements. Also, we conclude that any iodine cell filled at room temperature should be analyzed in the same manner in order to determine the temperature at which said jump occurs, which is highly dependent on the geometry of the cell and iodine content.\\
The outset of this study was the determination of a cell temperature at which the precision in radial velocity determination was the best, but since the achievable precision shows only a wavelength-dependency, this ``ideal'' operating temperature can be assessed by different factors. The transmission of the iodine cell is greatly reduced at temperatures $\gg300$~K and since the iodine content in this particular cell is relatively high, providing full wavelength coverage even at ambient temperatures, the operating temperature employed during stellar radial velocity measurements will be 300~K, a value well above the maximum ambient temperatures at night in the future observing location (Guanajuato, Mexico).\\